\documentclass[aip,jcp,reprint,floatfix,a4paper]{revtex4-1}

\usepackage{graphicx}
\usepackage[sort&compress]{natbib}
\usepackage[sumlimits]{amsmath}
\usepackage{amsfonts}
\usepackage{amssymb}
\usepackage{dcolumn}
\usepackage{subdepth}
\usepackage{color}
\usepackage{enumitem}
\usepackage{blindtext}
\usepackage{soul}
\usepackage{hyperref}
\usepackage{braket}

\newcommand{\HK}{HK}
\newcommand{\vV}{vV}
\newcommand{\Eh}{E_\mathrm{scl}^{\mathrm{H}}(t)}
\newcommand{\Ew}{E_\mathrm{scl}^{\mathrm{W}}(t)}

\newcommand{\Eq}[1]{Eq.~(\ref{#1})}
\newcommand{\Eqs}[1]{Eqs.~(#1)}
\newcommand{\Fig}[1]{Fig.~\ref{#1}}

\newcommand{\diff}{\mathrm{d}}
\renewcommand{\i}{\mathrm{i}}
\newcommand{\SI}{Supplement}
\begin{document}

\title{Semiclassical Propagation: Hilbert Space vs.\ Wigner Representation}

\author{Fabian Gottwald}
\affiliation{Institute of Physics, University of Rostock,
  Albert-Einstein-Str.\ 23-24, 18059 Rostock, Germany}
\author{Sergei D.\ Ivanov}
\email{sergei.ivanov@uni-rostock.de}
\affiliation{Institute of Physics, University of Rostock,
  Albert-Einstein-Str.\ 23-24, 18059 Rostock, Germany}

 \begin{abstract}
A unified viewpoint on the van Vleck and Herman-Kluk propagators in Hilbert space and their recently developed counterparts in Wigner representation is presented.
It is shown that the numerical protocol for the Herman-Kluk propagator, which contains the van Vleck one as a particular case, coincides in both representations.
The flexibility of the Wigner version in choosing the Gaussians' width for the underlying coherent states, being not bound to minimal uncertainty, is investigated numerically on prototypical potentials.
Exploiting this flexibility provides neither qualitative nor quantitative improvements.
Thus, the well-established Herman-Kluk propagator in Hilbert space remains the best choice to date given the large number of semiclassical developments and applications based on it.
\end{abstract}
\date{\today}

 \maketitle

\paragraph*{Introduction.}
The idea to utilize simple and robust tools of classical mechanics for describing quantum-mechanical effects is extremely tempting.
In this context, semiclassical techniques are attracting a never-fading interest, since van Vleck (\vV) proposed a semiclassical propagator which is exclusively based on classical trajectories.~\cite{VanVleck-PNAS-1928, Littlejohn1992}
However, the original semiclassical approaches suffer from conceptual intrinsic problems that mostly
originate from the fact that classical mechanics operates in phase space (position \textit{and} momentum), whereas a standard formulation of quantum mechanics is in Hilbert space (based on position \textit{or} momentum wavefunctions).
This makes a connection between the two theories not straightforward.
In particular, employing the original \vV\ formulation for numerical simulations is hindered due to the infamous root-search problem related to two-sided boundary conditions in position-space propagation.
This gave rise to initial-value representations (IVRs) that are exploiting initial position-momentum pairs instead of initial and final positions (or momenta), see Refs.~\onlinecite{Miller-JPCA-2001,Thoss-AR-2004,Kay-AR-2005} for reviews.
Further attempt to formulate the theory in terms of phase space objects was based on expanding the evolution operator in the coherent states' basis leading to the Herman-Kluk (\HK) propagator,~\cite{Herman-Kluk-CP-1984, Herman-Kluk-JCP-1986}
which can be
derived by different means~\cite{Kay-JCP-1994b, Kay2006,Grossmann-PLA-1998,Miller-MolPhys-2002,Miller-JPCB-2002,Child-JCP-2003,Deshpande-JPA-2006} and viewed as a frozen-Gaussian~\cite{Heller-JCP-1981} IVR approximation.

One can show that in their Hilbert space formulations the \vV\ propagator is a (non-optimal) limiting case of the \HK\ one if the width of the underlying Gaussians tends to zero either in positions or in momenta,
referred to as position \vV\ or momentum \vV, respectively.~\cite{Miller-JPCA-2001}
Interestingly, the \HK\ propagator can be also derived from the \vV\ propagator by either performing a stationary phase approximation~\cite{Grossmann-Book-2013} or by applying a modified Filinov transform.~\cite{Miller-MolPhys-2002}
These relations underline the principal semiclassical equivalence of the two propagators, although there have been misunderstandings about the \HK\ propagator being inferior to the \vV\ propagator, see Ref.~\onlinecite{Kay-AR-2005} for a detailed discussion.

A historically different way of putting classical and quantum mechanics on equal footing is to recast the entire quantum mechanics in terms of a phase-space quasi-probability density via, for instance, the Weyl transform.~\cite{DeAlmeida-PR-1998}
The resulting Wigner function constitutes a one-to-one representation of the quantum-mechanical density operator including coherences.~\cite{Wigner-PR-1932,Wigner-PRP-1984}
The latter are represented via small-scale oscillations,~\cite{Berry-PTRSA-1977,Zurek-NAT-2001} which 
pose a big challenge to any simulation technique and lead to strongly non-classical 
dynamics of Wigner functions even in the limit $\hbar \to 0$ due to ``dangerous cross-terms''.~\cite{Heller-JCP-1976}
These obstacles severely slowed down the development of the methodologies based on such a propagation.
Nevertheless, various semiclassical Wigner propagators have appeared in the past decade.~\cite{Dittrich-PRL-2006,Dittrich-JCP-2010,DeAlmeida2013}
An important contribution was made by Koda,~\cite{Koda-JCP-2015} based on the interpretation of the Moyal equation for the Wigner function,
as a Schr\"odinger equation.~\cite{Bondar-PRA-2013}
This approach provided a unified way to develop semiclassical Wigner propagators based on the existing Hilbert space ones.
Thereby, the Wigner version of the \HK\ propagator was suggested and the Wigner version of the \vV\ propagator~\cite{Dittrich-PRL-2006,Dittrich-JCP-2010} was re-derived and re-formulated in terms of an IVR.
The latter was again shown to turn into the former in the limit of  infinitely shrinked Gaussians, fully analogous to their Hilbert space counterparts.
As in Hilbert space, the \vV\ propagator constitutes a non-optimal particular case of the \HK\ propagator.

Practical applications of semiclassical techniques usually suffer from the so-called sign problem, which is caused by the rapid oscillations in the phase factors and leads to slow convergence of semiclassical averages. 
To tackle this problem, several approximations have been developed based on Filinov filtering~\cite{Walton-MolPhys-1996,Herman-CPL-1997} or forward-backward schemes~\cite{Sun-JCP-1999,Makri-CPL-1998} allowing one to deal with systems of up to hundreds of degrees of freedom (DOFs) in various contexts.~\cite{Kuehn-JPCA-1999,Batista-JCP-1999, Skinner-JCP-1999,Guallar-JCP-2000,Wang-JCP-2000,Ovchinnikov-JCP-2001,Nakayama-JCP-2003,Nakayama-CP-2004,Tao-JCP-2009,Herman-JPCB-2014,Alemi-JCP-2015}
Other approaches employ improved sampling techniques~\cite{Tao-JCP-2009,Tao-JCP-2014} or hybrid schemes treating a selected number of unimportant DOFs less accurately~\cite{Grossmann-JCP-2006,Buchholz-JCP-2016} or even implicitly as a heat bath.~\cite{Koch-PRL-2008,Koch-CP-2010}
The Hilbert space \HK\ propagator and the aforementioned approximations to it provide hitherto the most numerically convenient semiclassical protocols beyond the somewhat simplistic Wigner model also referred to as linearized semiclassical initial-value representation (LSC-IVR).~\cite{Miller-JCP-1998}
Thus, the natural question arises, whether the Wigner formulation of the \HK\ propagator has certain benefits over its Hilbert space counterpart.
This is the central question of this Communication.

\paragraph*{Theory.}
The Hilbert space formulation of the \HK\ propagator reads~\cite{Miller-JPCA-2001, Thoss-AR-2004,Kay-AR-2005} 
\begin{equation}
\label{eq: HK propagator Hilbert}
\exp\left[ -\frac{\i}{\hbar} \hat H t\right] \approx \intop \frac{\diff z_0}{2\pi\hbar} \,C_t(z_0) \exp\left[{\frac{\i}{\hbar} S_t(z_0)}\right] \ket{z_t}\bra{z_0}
\enspace ,
\end{equation}
where a one-dimensional system is considered for the sake of presentation;
the generalization to the 3D $N$-particle case is straightforward and is exploited in the \SI.
The integration here goes over the initial points of classical trajectories $z_t\equiv (q_t,p_t)$
on which coherent states $\ket{z}\equiv \ket{q,p}$ are placed.
The latter are defined via the wavefunctions
\begin{equation}
\label{eq: coherent state Hilbert}
\braket{x | z} \equiv  \left ( \frac{\gamma }{\pi \hbar}\right )^{1/4} \exp\left[-\frac{\gamma}{2 \hbar}(x-q)^2 + \frac{\i}{\hbar}p(x-q) \right]
\enspace .
\end{equation}
The parameter $\gamma>0$ balances the quantum uncertainties in position, $\sigma_q^2=\hbar/(2\gamma)$, and in momentum, $\sigma_p^2=\hbar \gamma/2$, keeping the uncertainty product minimal.
Further, the classical action, $S_t=\intop_0^t d\tau [\dot q_\tau p_\tau -H(z_\tau)] $, as well as the prefactor
\begin{equation}
C_t(z_0)= \left(   \frac{1}{2} \left[ \frac{\partial p_t}{\partial p_0}  + \frac{\partial q_t}{\partial q_0} -\i  \gamma \frac{\partial q_t}{\partial p_0}  + \frac{\i}{\gamma}\frac{\partial p_t}{\partial q_0} \right ] \right )^ {1/2}\,,
\end{equation}
containing the entries of the classical stability (monodromy) matrix appear in \Eq{eq: HK propagator Hilbert}.
As it was discussed above, the position/momentum IVR \vV\ propagator can be restored by considering the $\gamma \rightarrow \infty$/$\gamma \rightarrow 0$ limit, respectively.
Importantly, focusing on one canonical variable leads to an infinite uncertainty in the conjugate one.
This was even served as an advantage that may help to solve the longstanding problem of treating tunneling via semiclassical approaches.~\cite{Dittrich-JCP-2010}
Still, the absence of a physically meaningful, i.e.\ limiting distribution in one of the coordinates poses the question whether the respective phase space integral converges.
It is apparent that it can only converge by means of the fast-oscillating phase factor, which is at minimum hard to achieve numerically, as has been noted, e.g., by Kay.~\cite{Kay-JCP-1994}
Indeed, we confirm below that the \vV\ propagator neither solves the tunneling problem, nor performs well in general and in comparison to the \HK\ one in particular.
%
%It is worth mentioning that the \HK\ propagator does not have a correct asymptotic behaviour at $t \to 0$ due to ``spurious terms'' in the initial time derivatives;~\cite{Ankerhold-JCP-2002} interestingly, these terms vanish in the \vV\ limit ($\gamma \to \infty$).

Very recently, Koda has formulated a Wigner \HK\ propagator that, being applied
to the initial Wigner function, $W_0(z)$, reads~\cite{Koda-JCP-2015}
\begin{eqnarray}
\label{eq: HK phase space}
W(z,t) & = & \displaystyle \intop \frac{\diff \bar z_0 \, \diff \Delta z_0}{(2 \pi \hbar)^{2} } \, \tilde C_t(\bar z_0,\Delta z_0) \, \exp\left[\frac{\i}{\hbar} \tilde S_t(\bar z_0,\Delta z_0)\right] \nonumber \\
& \times & g(z ;\bar z_t, \Delta z_t) \intop \diff z' g^*(z'; \bar z_0, \Delta z_0) W_0( z')
\enspace .
\end{eqnarray}
Here, the integration is taken over initial centres (midpoints), $\bar z \equiv (z^+ + z^-)/2$,  and chords (differences), $\Delta z \equiv (z^+ -z^-)$, of pairs of classical trajectories, $z_t^\pm$. 
These carry 'phase space coherent states' defined as
\begin{eqnarray}
g(z;\bar z, \Delta z) & = & \det \left ( \frac{\Gamma}{\pi \hbar } \right )^{1/4} \exp\left[ - \frac{1}{2 \hbar} (z-\bar z)^T \Gamma (z- \bar z)\right]  \nonumber \\ 
& & \times \exp\left[ \frac{\i}{\hbar}  \Delta z^T \mathbf J^T (z-\bar z)\right] 
\enspace,
\end{eqnarray}
which can be viewed as operating on a double phase space.
Here centres and chords have the same conjugate relation as positions and momenta in the Hilbert-space coherent states.~\cite{Koda-JCP-2015}
The symmetric and positive definite ($2 \times 2$)-matrix $\Gamma$ balances the uncertainties in centres and chords and $\mathbf J$ is the symplectic matrix
\begin{equation}
 \mathbf J=
 \begin{pmatrix}
   0 & 1\\
  -1 & 0\\
 \end{pmatrix}
 \enspace .
\end{equation}
The action in the phase factor is defined as $\tilde S_t \equiv \intop_0^t \diff \tau  \left [ \dot {\bar z}_\tau ^T \mathbf J \Delta z_\tau - H^+(z^+_\tau) + H^-(z^-_\tau) \right ]$, where
the trajectories are evolved according to not necessarily the same Hamilton functions, $H^\pm$, 
and the prefactor $\tilde C_t$ consists of the respective stability matrices $\mathbf M_t^\pm$
\begin{eqnarray}
\label{eq: prefactor phase space}
\tilde C_t(\bar z_0, \Delta z_0) & = &(\det \Gamma)^{-1/2}   \det \left[ \frac{1}{2}\left ( \frac{ \Gamma}{2} + \i \mathbf J \right ) \mathbf M_t^+ \left ( 1+  \i \mathbf J  \frac{ \Gamma}{2} \right) \right.\nonumber \\ 
 &+&  \left. \frac{1}{2} \left ( \frac{ \Gamma}{2} - \i \mathbf J \right) \mathbf M^-_t \left ( 1-  \i \mathbf J  \frac{ \Gamma}{2}   \right)  \right]^{1/2}
\enspace .
\end{eqnarray}
As in Hilbert space, it is possible to consider the \vV\ limits
%SI: In principle it is better to write det explicitly.
$|\Gamma| \rightarrow \infty$ and $|\Gamma| \rightarrow 0$ leading to pure 'centre' and  'chord' representations, respectively.~\cite{Koda-JCP-2015}
Hence, we refer to these limits as centre \vV\ and chord \vV\ propagators.

\begin{figure}[tb]
	\includegraphics[width=0.70\columnwidth]{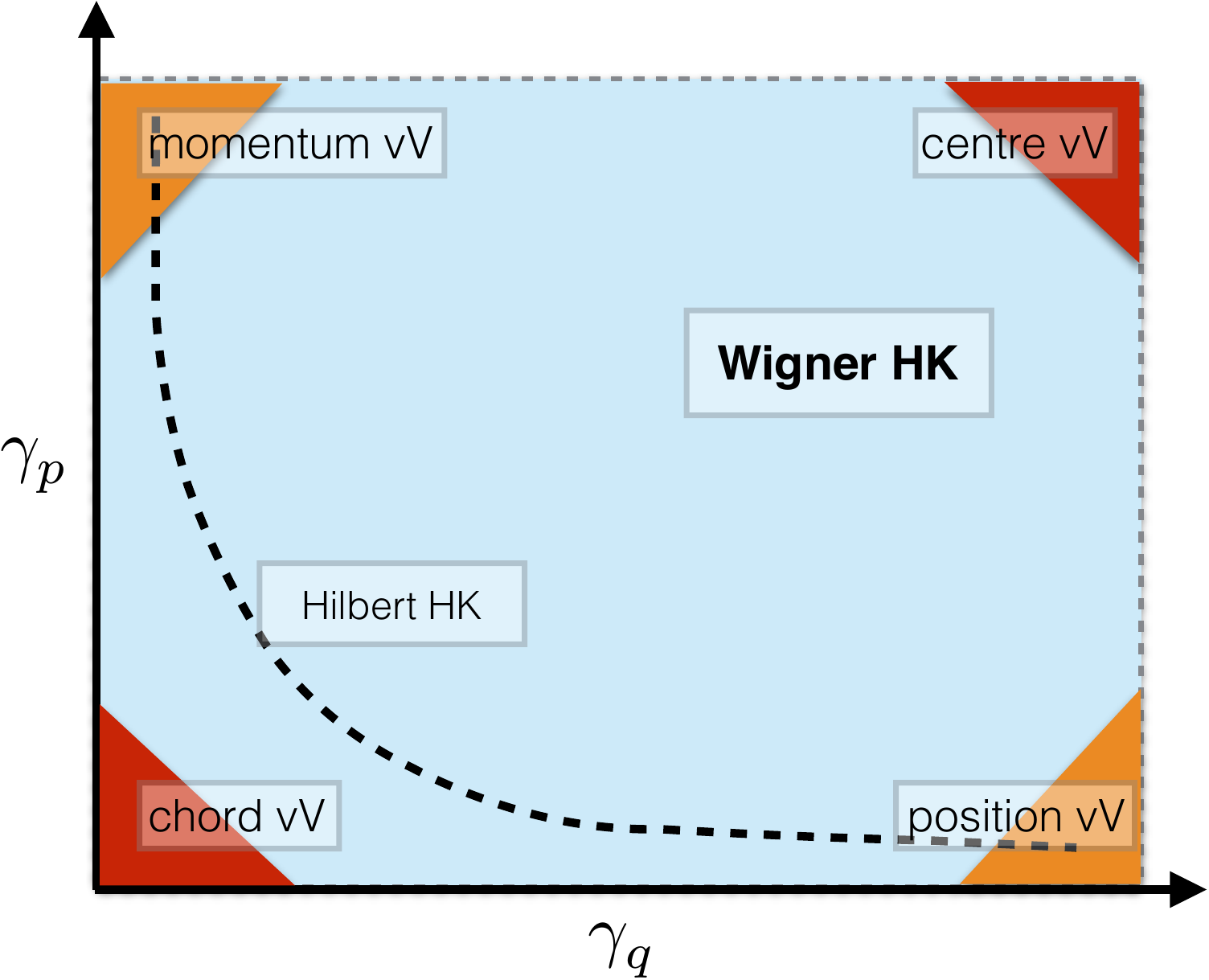}
	\caption{
		\label{fig: sketch}
		Schematic relationship of all semiclassical IVR propagators, see text for the discussion.}
\end{figure}

All aforementioned propagators are schematically depicted in \Fig{fig: sketch}.
Assuming the form $\Gamma=\mathrm{diag}(\gamma_q, \gamma_p)$, the Wigner \HK\ propagator covers the whole $(\gamma_q,\gamma_p)$ - plane.
Importantly, Weyl-transforming the density operator propagated via two Hilbert space \HK\ propagators, \Eq{eq: HK propagator Hilbert}, leads to a propagator that has the same form as the Wigner \HK\ propagator in \Eq{eq: HK phase space} but inherits the minimal uncertainty condition,~\cite{Koda-JCP-2015}
which can be written as $\gamma_q \cdot \gamma_p=4$ and is represented by the blue dotted hyperbola in \Fig{fig: sketch}.
The Hilbert space position- and momentum \vV\ limits introduced above are characterised by $(\gamma_q, \gamma_p) \rightarrow (\infty, 0)$ and $(\gamma_q, \gamma_p) \rightarrow (0, \infty)$ and are located in the lower right and upper left corners of \Fig{fig: sketch}, respectively.
These limits preserve the minimal-uncertainty condition and are thus accessible via the Hilbert space \HK\ propagator.
In contrast, the centre- and chord \vV\ limits are given by $(\gamma_q, \gamma_p) \rightarrow (\infty, \infty)$ and $(\gamma_q, \gamma_p) \rightarrow (0, 0)$ located in the upper right and lower left corner.
They can not be reached by the Hilbert space \HK\ propagator as they disobey the minimal uncertainty condition.
Overall, \Fig{fig: sketch} illustrates that the Wigner \HK\ propagator covers all important semiclassical IVR propagators that have been formulated so far.

\paragraph*{Theoretical considerations.}
The relationships in \Fig{fig: sketch} can only be uncovered by a direct transform of the Hilbert space propagators into the Wigner representation.
Without this explicit transform, it is natural to expect that the two versions of the \HK\ propagator might lead to conceptually different numerical schemes.
Let us consider a general expectation value, $E(t)$, of the form
\begin{equation}
\label{eq: expectation value Hilbert}
E(t)= \mathrm{Tr} \left ( \hat A \exp \left[-\frac{\i}{\hbar} \hat H^{+} t \right] \hat B \exp\left[\frac{\i}{\hbar} \hat H^{-} t \right]\right )
\enspace ,
\end{equation}
with two Hermitian operators $\hat A$, $\hat B$ and  two time-evolution operators with respect to the Hamiltonians $\hat H^\pm$.
With $\hat H^+=\hat H^-$ this expression covers standard quantum expectation values and correlation functions of an observable $\hat A$ by identifying $\hat B$ with the density operator $\hat \rho$
or with $\hat \rho \cdot  \hat A$, respectively.
Further, \Eq{eq: expectation value Hilbert} can account for non-Heisenberg time evolution with $\hat H^+\neq\hat H^-$, which appears, e.g., in optical response functions for electronic transitions.
The semiclassical expression in Hilbert space, $\Eh$, can be obtained straightforwardly by inserting two Hilbert space \HK\ propagators, \Eq{eq: HK propagator Hilbert}, into \Eq{eq: expectation value Hilbert}, resulting in 
 \begin{equation}
 \label{eq: scl expectation Hilbert space}
 \Eh=\intop \frac{\diff z^{+}_0 \diff z^{-}_0}{(2\pi \hbar)^{2}} G_t(z^+_0,z^-_0) A^*(z^+_t, z^-_t) B(z^+_0,z^-_0)
  \end{equation}
 with
 \begin{eqnarray}
 \label{eq: functions Hilbert space}
 G_t(z^+_0,z^-_0) & \equiv & C_t(z^+_0) C^*_t(z^-_0) \exp\left[\frac{\i}{\hbar} \{S_t(z_0^+)-S_t(z_0^-)\}\right] \nonumber \\
 A(z_t^+,z_t^-) & \equiv & \braket{z_t^+ | \hat A | z_t^-} \nonumber \\
 B(z_0^+,z_0^-) & \equiv & \braket{z_0^+ | \hat B | z_0^-} 
 \enspace .
 \end{eqnarray}
In the Wigner representation, $E(t)$ reads
\begin{equation}
\label{eq: expectation value Wigner}
E(t)=\intop \diff z\,A_\mathrm{W}(z) B_\mathrm{W}(z,t)
\enspace ,
\end{equation}
where $A_\mathrm{W}(z)$ and $B_\mathrm{W}(z,t)$ are the Weyl symbols representing the operators $\hat A$ and $\exp[-\frac{\i}{\hbar} \hat H^{+} t ] \hat B \exp[\frac{\i}{\hbar} \hat H^{-} t]$, respectively.~\cite{DeAlmeida-PR-1998}
The corresponding semiclassical expression, $\Ew$, follows directly from applying \Eq{eq: HK phase space} to the Weyl symbol, $B_\mathrm{W}(z,t)$, in \Eq{eq: expectation value Wigner} and leads to
\begin{equation}
 \label{eq: scl expectation phase space}
 \Ew\!=\!\!\intop \frac{\diff \bar z_0 \diff \Delta z_0}{(2 \pi \hbar)^{2}} \tilde G_t(\bar z_0, \Delta z_0) \tilde A_\mathrm{W}(\bar z_t, \Delta z_t) \tilde B_\mathrm{W}(\bar z_0, \Delta z_0) \,,
 \end{equation} 
with the functions
\begin{eqnarray}
\label{eq: functions phase space}
\tilde G_t(\bar z_0, \Delta z_0) & \equiv & \tilde C_t(\bar z_0, \Delta z_0) \exp\left [\frac{\i}{\hbar} \tilde S_t(\bar z_0, \Delta z_0)\right] \nonumber \\
\tilde  A_\mathrm{W}(\bar z_t, \Delta z_t) & \equiv & \intop \diff z A_\mathrm{W}(z) g(z;\bar z_t, \Delta z_t) \nonumber \\
\tilde B_\mathrm{W}(\bar z_0, \Delta z_0) & \equiv & \intop \diff z B_\mathrm{W}(z) g^*(z;\bar z_0, \Delta z_0)  \,.
\end{eqnarray}
Although the expressions for $\Eh$ and $\Ew$ look similar, they seem to contain different ingredients thus leading to different computation protocols.
However, a careful rearrangement of all the functions in \Eqs{\ref{eq: functions Hilbert space}, \ref{eq: functions phase space}} using the transform from initial centres and chords to initial points of the trajectories, carried out in detail in the \SI, shows that the integrands of
$\Eh$ and $\Ew$ coincide exactly for a matrix $\Gamma$ satisfying, again, the minimal uncertainty condition $\gamma_q \cdot \gamma_p=4$.
Importantly, this coincidence can be shown \textit{without} the initial Weyl transform and one can, thus, conclude that the semiclassical Wigner propagator does not yield a principally new computation protocol compared to its Hilbert space version.
This is the first main result of this Communication.

\begin{figure}[tb!]
 \centering
 %\begin{widetext}
 % \begin{minipage}{\textwidth}
      \includegraphics[width=0.99 \columnwidth]{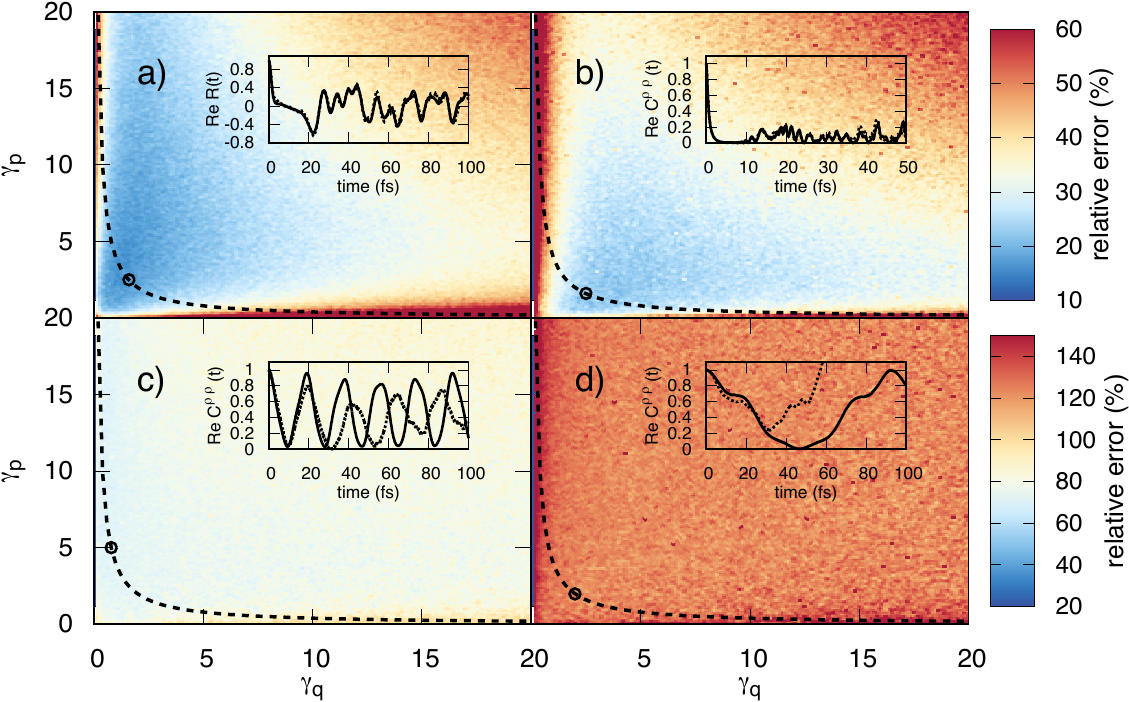}
	\caption{\label{fig: error}
The error, \Eq{eq:error}, displayed as a function of $\gamma_q$ and $\gamma_p$ for the optical response function of two shifted Morse potentials (panel a) and the state-state auto-correlation function for the double-well potential in the classical, shallow, and deep tunneling regimes (panels b-d), see text. 
Insets compare a typical semiclassical function in question (dotted line) against the exact result (solid line).
The parameters $(\gamma_q,\gamma_p)$ corresponding to former are marked via the black circle in the 2D error plots. 
Dashed line indicate the minimal uncertainty hyperbola.
}
%  \end{minipage}
 %\end{widetext}
\end{figure}

\paragraph*{Numerical investigations.}
%FG: We can put this sentence also as a second conclusive sentence in the section before!
%SI: Perhaps we do not have to. There we talked about _principal_ novelty.
According to the theoretical considerations above, the only new aspect of the Wigner \HK\ propagator is a mere technical flexibility in choosing the width matrix $\Gamma$ being not bound to minimal uncertainty.
It remains to be checked whether this flexibility can yield a numerical advantage,
which is done below based on two prototypical examples. 
First, the optical response function, $R(t)$, of an electronic transition between two shifted potential energy surfaces (PESs), $V_{\pm}(q)$, is investigated.
Such a response function is a particular case of \Eq{eq: expectation value Hilbert}, with $\hat A=1$ and $\hat B=\hat \rho_0$, i.e.~the initial state, and, being Fourier-transformed, corresponds to a vibronic spectrum in the Condon approximation.
The two Hamiltonians, $\hat H^{\pm}$, describe the nuclear time evolution on
the respective PESs, $V^{\pm}(q)$, which have the form of Morse potentials.
In order to test a chemically relevant regime, $V^+(q)$ is chosen to be the ground state PES of the O$-$H stretch in water, with parameters taken from Ref.~\onlinecite{Paesani-JCP-2010}, whereas those for $V^-(q)$ are set to yield half of the harmonic frequency of $V^+(q)$.
The
%SI: In fact:
upper PES is
% two PESs are 
shifted by 0.2\,\AA\ and the mass of the nuclear DOF, $q$, is the reduced mass of oxygen and hydrogen, see \SI\ for further details.
%
%SI: Shouldn't it be mensioned that it is fit to a Gaussian?
The ground state of $V^+(q)$ in the Wigner representation is used as the initial state.

Second, we consider a proton in the double-well potential from Ref.~\onlinecite{Borgis1992}.
This potential describes solvent-induced proton transfer in strongly H-bonded complexes and has a barrier height of 4\,kcal/mol, which amounts to two quantum eigenstates below the barrier only.
For this system, the state-state auto-correlation function, $C^{\rho\rho}(t)$,
is calculated, which can be obtained from \Eq{eq: expectation value Hilbert} by setting $\hat A=\hat B=\hat \rho_0$
 and $\hat H^+ =\hat H^-$.
A Gaussian Wigner function, corresponding to the part of the first excited state localised in one of the wells, is taken as the initial state.
Schematically, one can identify three characteristic regimes of tunneling: 
%FG: Again, I have to repeat my question: isn't 'classical' misleading? This would imply that everything behaves classically which is not true even here!
%SI: I think it is clear, what is meant. If you do not like the term, then suggest something else.
i)~classical tunneling, where the classical trajectories can easily pass the barrier,
ii)~shallow tunneling,
a regime where tunneling is less important but still plays a role (see also Ref.~\onlinecite{Tanaka-PRA-2006}) and
iii)~deep tunneling regime. 
%
%SI: Wouldn't it be logical to swap the order below as well?
In the deep tunneling regime, the chosen initial state is centred at the minimum of one of the wells (average energy of 88\,\% of the barrier height) and
%SI: Can be viewed or corresponds to?
%FG: It doesn't strictly corresponds to it (there are also higher state mixed in). I would keep 'viewed as' because it is rather a model picture.
can be viewed as a superposition of the two eigenstates below the barrier.
This leads to a slow Rabi-type oscillation of the probability density between the wells, which is mainly induced by tunneling.
This characteristic motion is reduced when the energy of the initial state is increased, e.g., upon placing it away from the potential minimum. 
In the shallow tunneling regime the initial state is, thus, centred near the barrier top (130\,\% of the barrier energy) and strongly displaced from the minimum (820\% of the barrier energy) in the classical regime.
%
%SI: Is it an old TODO?
%FG: I'm not completely satisfied with the shallow and deep tunneling regimes though. To discuss (TODO)

In order to quantify the accuracy of the semiclassical treatment, the expectation values, $E_\mathrm{scl}(t)$, have been computed via \Eq{eq: scl expectation phase space}
and compared against exact quantum references, $E_\mathrm{qm}(t)$, obtained from a second-order split-operator treatment of the Wigner propagation.
We have employed 25,000 and 50,000 trajectory pairs 
%to achieve statistical convergence
for the first and the second example, respectively, see \SI\ for further numerical details.
Note that the Hilbert space \HK\ propagator protocol is covered by \Eq{eq: scl expectation phase space} with $\Gamma$ fulfilling the minimal uncertainty condition.
The error in the semiclassical result has been quantified as
\begin{equation}
\label{eq:error}
\epsilon(\gamma_q,\gamma_p) \equiv \sqrt{\frac{\intop \diff t |E_\mathrm{qm}(t)-E_\mathrm{scl}(t)|^2}{\intop \diff t |E_\mathrm{qm}(t)|^2}}
\enspace ,
\end{equation}
being a function of the two parameters, $\gamma_q$ and $\gamma_p$, for a diagonal matrix $\Gamma = \mathrm{diag}(\gamma_q,\gamma_p)$.

The error for the first example, the optical response function, is displayed in \Fig{fig: error}a), as a 2D 
colour plot; note that each panel has the same layout as \Fig{fig: sketch}. 
The minimal uncertainty hyperbola, that corresponds to the Hilbert space \HK\ propagator, is indicated via the dashed line therein.
%
%SI: plateua is wrong term! It means something that is high and planar.
The error reveals a shallow minimum ($\epsilon \approx 12\,\%$).
%a plateau (blue area) where the error is minimal ($\approx12\%$).
%
In the inset, a representative response function, $R_\mathrm{scl}(t)$ (dotted line), calculated at the position indicated by a circle, is compared against the reference ,$R_\mathrm{qm}(t)$, (solid line) demonstrating that this error corresponds to a visually good agreement. 
Importantly, this 
%SI: 
shallow minimum
%plateau of minimal errors
is crossed by the minimal uncertainty line, implying that it can be accessed by the Hilbert space propagator. 
%It can be also seen that the optimal choice is $\gamma_p \approx \gamma_q$.
%
In contrast, the \vV\ limits corresponding to the corners of the graph, see \Fig{fig: sketch}, reveal large errors confirming their numerically poor performance.
It is though expected that a dramatic increase in the number of trajectory pairs will eventually yield accurate results.

Considering the accuracy of the state-state autocorrelation function of the double-well system, panels b) - d) in \Fig{fig: error}, one observes the very same behaviour for the classical tunneling regime, panel b): a 
%SI:
shallow minimum
%plateau of minimal errors 
($\epsilon \approx 20\,\%$) is clearly visible and accessible by the minimal uncertainty line. 
The \vV\ limits, in turn, show large deviations from the exact reference.
The reasonable accuracy in this regime is expected, since tunneling does not play a role as the system has enough energy to pass the barrier classically.
In contrast, in the shallow and deep tunneling regimes, panels c) and d), a bad accuracy of the semiclassical approximation is revealed.
While the minimum of the error (with large errors of $\approx 65\,\%$) can still be recognized in the shallow tunneling regime (around the circle in panel c)), the deep tunneling regime is not accurately described on the whole $(\gamma_q,\gamma_p)$-plane.
This is also reflected in a direct comparison of the semiclassical result, $C^{\rho\rho}_\mathrm{scl}(t)$, against the exact reference, $C^{\rho\rho}_\mathrm{qm}(t)$, see insets.
Thus, the infamous breakdown of the semiclassical approximation in the deep tunneling regime~\cite{Grossmann-CPL-1995,Kay1997} is not prevented by exploiting the full flexibility in choosing the matrix~$\Gamma$ and the performance is even worse in the \vV\ limits.
Overall, the numerical investigation shows that whenever the semiclassical results are accurate, the shallow error minimum is accessible via the minimal uncertainty line, i.e.\ via the Hilbert space \HK\ propagator.
If the semiclassical performance is bad, utilizing the broader parameter range of $\Gamma$ provided by the Wigner \HK\ propagator does not yield an improvement.

\paragraph*{Conclusions.}
In this Communication, we have compared semiclassical Herman-Kluk and van Vleck propagators formulated in Hilbert space and in Wigner representation.
It has been shown that switching to the Wigner representation yields the very same numerical protocol as in Hilbert space and is thus not beneficial.
The only additional flexibility of the Wigner formulation lies in the arbitrary choice of the Gaussians' width for the underlying coherent states being not bound to minimal uncertainty.
After a careful numerical investigation for prototypical Morse and double-well potentials it has turned out that this freedom leads neither to qualitative nor quantitative improvements in performance.
The overall conclusion is thus that the semiclassical Herman-Kluk propagator in Wigner representation 
performs exactly the same way as its state-of-the-art counterpart in Hilbert space. 

The authors would like to thank Thomas Dittrich, Frank Grossmann and Oliver K\"uhn for fruitful and stimulating discussions.
S.D.I.\ gratefully acknowledges the financial support by the Deutsche Forschungsgemeinschaft (IV~171/2-1). 

\paragraph*{Supplementary Material}

The supplementary material contains the detailed proof that the Wigner and Hilbert-space versions of the \HK\ propagator yield the same protocol for computing semiclassical averages.
It further gives a detailed system description and elaborates on simulation strategies employed.

\bibliography{./semiclassics}
\end{document}